\documentclass[conference]{IEEEtran}
\IEEEoverridecommandlockouts
\usepackage{cite}
\usepackage{amsmath,amssymb,amsfonts}
\usepackage{algorithmic}
\usepackage{graphicx}
\usepackage{textcomp}
\usepackage{xcolor}
\usepackage{url}
\def\BibTeX{{\rm B\kern-.05em{\sc i\kern-.025em b}\kern-.08em
    T\kern-.1667em\lower.7ex\hbox{E}\kern-.125emX}}
\begin{document}

\title{Prompt Optimization for User Simulation in Conversational Recommender Systems: A Multi-Objective Framework\\
}

\author{\IEEEauthorblockN{Nipun B Nair}
\IEEEauthorblockA{
\textit{Monash University, Australia}\\
nipun.nair@monash.edu}
\and
\IEEEauthorblockN{Tongtong Wu}
\IEEEauthorblockA{
\textit{Monash University, Australia}\\
tongtong.wu@monash.edu}
\and
\IEEEauthorblockN{Weiqing Wang}
\IEEEauthorblockA{
\textit{Monash University, Australia}\\
teresa.wang@monash.edu}
}
\maketitle

\begin{abstract}
Conversational recommender systems (CRSs) are a core component of next-generation intelligent recommender systems because they enable users to actively elicit preferences, clarify intentions, and adapt recommendations in real time. However, there are two key obstacles in the CRS domain: evaluation and access to training data. Evaluating CRSs through real human studies is more critical than for traditional recommender systems, yet such studies are both costly and time-consuming. Moreover, CRS interaction data are often difficult to obtain for model training due to privacy concerns. Large language model (LLM)-based user simulators have shown promise in addressing both challenges by generating synthetic user interactions for evaluation and training. However, existing approaches suffer from systematic positive bias, data leakage, and limited behavioral diversity, and they rely on brittle manual prompt engineering that requires extensive domain expertise. In this paper, we propose a framework to automatically optimize prompts for LLM-based user simulators in CRSs, simultaneously mitigating these issues. Experimental results demonstrate that the proposed framework achieves improved behavioral alignment with human interaction patterns compared to baseline methods across diverse prompt settings.
\end{abstract}

\begin{IEEEkeywords}
User simulation, Prompt tuning, Conversational Recommendation Systems, LLMs
\end{IEEEkeywords}

\section{Introduction}

Recommender systems \cite{Fan2024RSinEraOfLLMs,Lin2023RSBenefitFromLLMsSurvey} play a critical role in information seeking by enabling users to efficiently discover relevant items, content, and information. These systems deliver both user value and business impact and  continue to drive strong academic and industrial interest \cite{Zhang2023RecommendationAsInstruction, Fan2024RSinEraOfLLMs,Wu2024OptimizingER,RecDCL}. Conversational recommender systems (CRSs) enhance this capability by engaging users in natural language to infer user preferences and reasons behind those preferences. CRSs allows users to articulate preferences, explore options interactively, and provide fine-grained feedback \cite{Gao2021CRSSurvey,Zhang2020EvalCRSUserSimulation} which is fundamentally different from traditional recommender systems and have been recognized as the core component of next-generation intelligent recommender systems in this Large Language Model (LLM) era\cite{Li2018DeepConversationalRecommendations,wang2024recmind,Gao2021CRSSurvey}. 

Conversational recommender systems typically require extensive user testing prior to deployment, creating costly and time-consuming bottlenecks in industrial development pipelines, as evaluation depends on real users interacting with the system. Additionally, collecting conversational data introduces privacy concerns that hinder deployment in regulated settings (e.g., healthcare and financial area). User simulators provide a promising solution to these challenges by enabling scalable, low-cost, and privacy-preserving evaluation and training of CRSs without requiring extensive interaction with real users. 

User simulators are automated agents that emulate human interaction patterns in recommendation dialogues by generating responses that approximate real user behavior. LLM  based user simulators have shown its great potential in this LLM era by serving as synthetic user simulators that enable reproducible, cost-effective user simulation in recommender systems across diverse user populations and interaction contexts
\cite{balog2025user,Wang2024UserBehaviorSimulation,Bernadic2025LLMsPowerSystems,przystalski2025rise,xiang2024simuser}. 
 
The existing LLM user simulators can be categorized into fine-tuned/training based simulators\cite{chen2025recusersim, Zhao2021UserSim, Zhang2025LLMPoweredUserSimulator} and prompt based simulators \cite{Yoon2024LLMsAsUserSimulators, zhu2024reliable, zhu2025llm, luo2024duetsim, Chen2025HumanFreePromptTuning}. In this paper, we focus on prompt-based simulators, as they are cost-effective compared to training/fine-tuning based methods which require fine-tuning of a large language model. LLM-based user simulators operate by taking a structured user profile and a recommended item as input, then generating a natural language response that approximates real user behavior. In prompt-based simulators, this behavior is controlled entirely through instructions provided to a frozen LLM, without modifying model weights, contrasting with fine-tuned approaches that require large labeled datasets and significant compute, making them impractical in privacy-sensitive or resource-constrained settings. 

Current prompt-based LLM user simulators face three critical challenges that limit practical deployment in conversational recommender systems:

\textbf{(1) Systematic positive bias:} LLM-based simulators exhibit unrealistically high acceptance rates, generating overwhelmingly positive responses that fail to capture realistic rejection behavior. For instance, when presented with movie recommendations, LLM-based simulators tend to accept most of suggestions regardless of stated user preferences, while real users reject nearly half of the presented recommendations \cite{Yoon2024LLMsAsUserSimulators, Zhang2020EvalCRSUserSimulation}. 

\textbf{(2) Data leakage:} Unlike standard train-test leakage in machine learning, data leakage in user simulation occurs when user's profile history items appear as item mentions verbatim. Rather than reasoning from genuine preferences, the simulator echoes these items back in its responses, inflating apparent accuracy and artificially suppressing diversity metrics in ways that misrepresent true simulator capability \cite{zhu2024reliable, Yoon2024LLMsAsUserSimulators}.

\textbf{(3) Severely constrained behavioral diversity:} Simulators exhibit popularity bias, disproportionately favoring well-known mainstream items over niche content, and temporal clustering, concentrating recommendations within narrow time windows. This leads to homogenized item mentions and fails to capture the diverse preferences of real user populations, limiting their effectiveness for evaluating recommender systems across varied user personas\cite{Zhang2020EvalCRSUserSimulation,Zhao2021UserSim,wang2025limits}. 

While prompt-based simulators are more accessible, their performance is highly sensitive to prompt quality \cite{zhan-etal-2023-user, Lester2021PromptTuningScale, PromptMM}, and recent frameworks addressing this lack comprehensive solutions that tackle all three challenges simultaneously \cite{zhu2025llm, chen2025recusersim, luo2024duetsim}. A critical gap therefore remains in principled methods for automatically optimizing simulator behavior to achieve realistic behavioral patterns. Beyond CRS evaluation, this work addresses a core challenge in LLM-driven data engineering: how to automatically generate, optimize, and validate synthetic behavioral data at scale.
 
To solve these issues, we propose a framework which achieves joint optimization of simulating user profiles and response patterns and overcomes the three challenges simultaneously. Moreover, the framework also avoids heavy manual prompt engineering through automatic prompt optimization via interpretable text based gradients. In this paper we make the following contributions. 
\begin{itemize}
    \item \textbf{Proposal of optimized framework:} To the best of our knowledge, we are the first one proposing a framework to automatically optimize the prompt for LLMs based user simulator in CRSs by overcoming the three challenges simultaneously.
    \item \textbf{Solution of Challenges:} In the framework, we design entropy-aware and textual-gradient-based scoring functions that solve the problems of positive bias and popularity bias. Additionally we design a profile summarization way to reduce the impact of data leakage while retaining the essential profile information. To evaluate the simulator's ability in overcoming overacceptance bias, we propose NegFeedback, a proof-of-concept metric specifically designed to assess the correctness and rationale of negative user feedback.
    \item \textbf{Improved behaviour alignment:} We demonstrate improved behavioral alignment with human interaction patterns compared to GPT-3.5 and GPT-4 baselines, using local execution with Llama3.3. We further validate the proposed metric NegFeedback using both an LLM-based evaluator and human evaluators. The LLM evaluator enables us to scale the evaluation, and its alignment with human judgments demonstrates the validity of the LLM evaluator.
\end{itemize}

\section{Related Work}
In this section, we organize our review around key challenges to establish the foundation for our approach.
\subsection{LLM-Based User Simulation in Recommender Systems}
 Recent implementations demonstrate significant advances in generating human like responses and maintaining coherent user personas. Many studies have explored LLM powered simulator adapted for recommendation contexts that leverages contextual understanding and incorporate personality traits and demographic features into user modeling to generate more realistic user interactions\cite{Wang2024UserBehaviorSimulation,Zhang2025LLMPoweredUserSimulator,Park2023GenerativeAgents,Zhang2024GenerativeAgentsInRecommendation}. However, existing implementations rely predominantly on manual prompt engineering, where researchers craft static templates based on intuition and limited experimentation \cite{Yoon2024LLMsAsUserSimulators,zhan-etal-2023-user}. This manual approach leads to several fundamental problems: prompt brittleness across different domains, inability to systematically optimize for real user behavior, and lack of principled methods for incorporating domain specific requirements. Recent comprehensive surveys \cite{Wu2023LLMsForRecommendationSurvey,Lin2023RSBenefitFromLLMsSurvey} acknowledge these limitations but do not provide systematic solutions for prompt optimization in user simulation contexts. The need for automatic prompt optimization in user simulation is further highlighted by the scalability challenges of manual approaches. As recommendation systems become more complex and diverse, manually crafting prompts for each domain and user type becomes increasingly impractical. 

\subsection{Prompt Optimization and Engineering for LLMs}
Recent Automated Prompt Optimization (APO) methods include gradient-based, evolutionary, and RL-based approaches\cite{li2025survey,ramnath2025systematic}. TextGrad \cite{yuksekgonul2024textgrad} represents a significant advancement by optimizing prompts through iterative refinement based on textual feedback from the target LLM itself. This black box approach offers key advantages: (1) supports model optimization without requiring model access, and (2) allows for dynamic domain specific adaptation. However, existing APO methods have not addressed user simulation challenges. Current APO method applications focus on traditional NLP tasks with well-defined success metrics, while user simulation requires handling conflicting behavioral objectives, complex behavioral patterns beyond simple accuracy, and authentic human behavioral diversity. 

\subsection{Evaluation Metrics and Feedback Mechanisms}
Existing user simulation evaluations rely on surface-level NLP metrics (e.g., BLEU, F1) that overlook behavioral fidelity, such as realistic acceptance rates and informative rejections\cite{zhu2024reliable,zhang2024usimagent,chen2025recusersim}. While recent work acknowledges these gaps, it lacks systematic methods to assess rejection quality or alignment with human behavior\cite{zhu2025llm,chen2025recusersim}. These gaps motivate our development of NegFeedback, a proof-of-concept metric designed to assess authenticity and rationale quality of user rejection in user simulation contexts.

\subsection{The Bias and Leakage Problem in LLM-Based User Simulation}

LLM-based simulators suffer from systematic positive bias leading to unrealistically high acceptance rates and popularity skewed responses \cite{Sekulic2024AnalysingUtterances,Gui2023ChallengeLLMSimBehavior,Lu2024LLMsABMComplexSystems}. Prompt sensitivity exacerbates these issues, while data leakage allows access to target items, compromising evaluation integrity\cite{zhu2024reliable}. Existing methods lack principled tools for aligning simulator behavior with human patterns. Our framework extends automatic prompt-tuning with bias correction to address these challenges and improve behavioral fidelity.
\begin{figure}[htbp]
\centering
\includegraphics[
  width=0.5\textwidth,
  height=0.8\textheight,
  keepaspectratio]{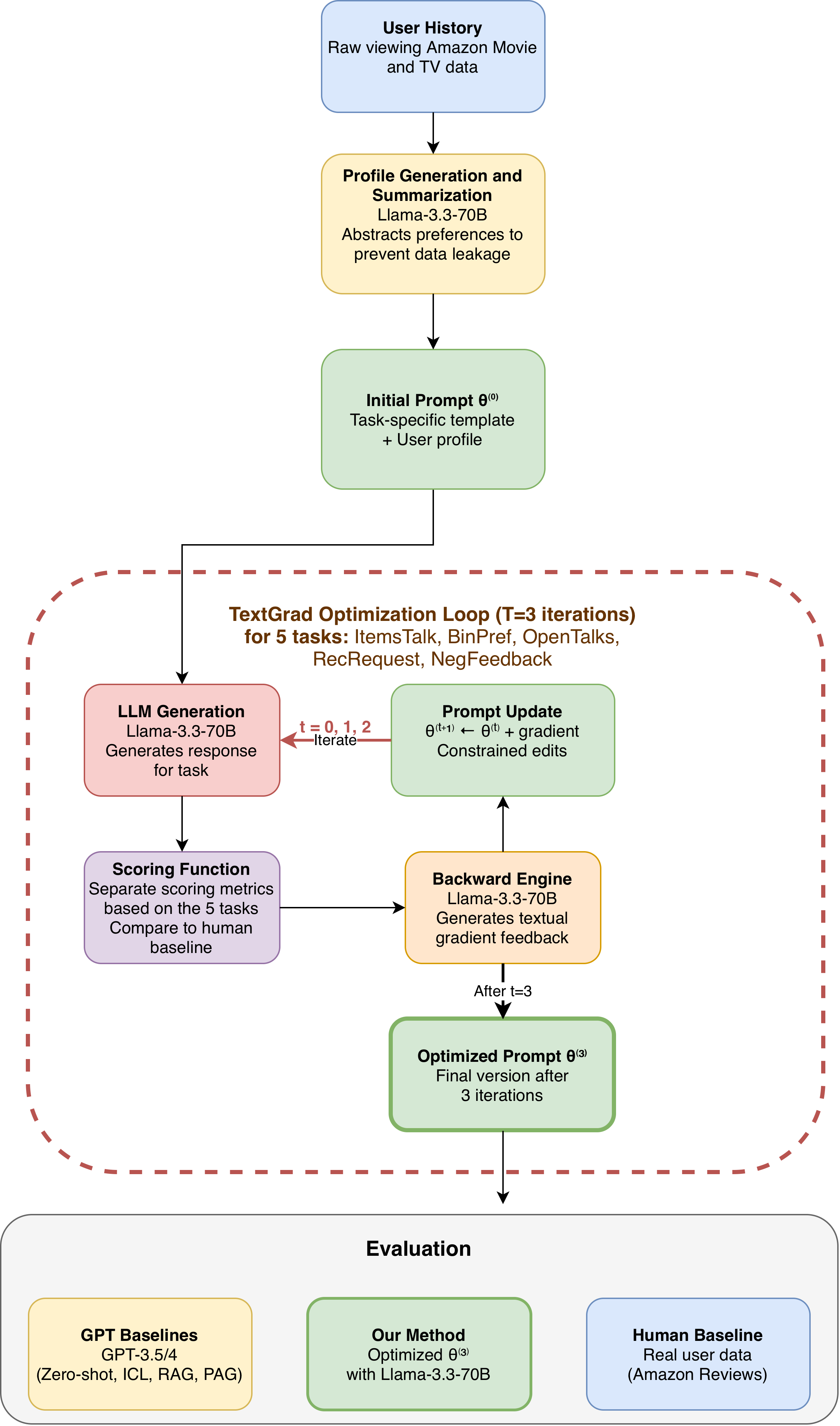}
\caption{Automatic Prompt tuning-based user simulator optimization pipeline}
\label{fig:framework}
\end{figure}
\section{Our Framework}
\label{sec:our_framework}
Compared with existing LLM based user simulators in CRSs, our framework leverages TextGrad \cite{yuksekgonul2024textgrad} to achieve the automatic prompt optimization to avoid the heavy prompt engineering. To effectively adapt to user simulation task in CRSs,  our framework optimizes behavioral fidelity rather than linguistic similarity. Moreover, our framework introduces an entropy-aware and textual-gradient-based scoring functions to teach the model to optimize prompts automatically to overcome the positive bias and popularity bias. Moreover, we also design a profile summarization approach to achieve a good trade-off between reduced data leakage risks and keeping essential profile information. Lastly, a newly  designed metric which will be carefully validated is proposed to evaluate the simulators' rejection quality to measure their ability in overcoming over-acceptance bias. The framework is based on a local LLM Ollama's Llama-3.3-70B for privacy-preserving execution.

Figure \ref{fig:framework} illustrates the overall pipeline of our approach.
The process starts with raw user interaction histories extracted from the Amazon dataset (Section \ref{subsec:dataset}). These histories are then used to generate and summarize user profiles (Section \ref{subsec:profile_summary_for_data}). Based on the resulting profiles, task-specific initial evaluation prompts are constructed (Section \ref{subsec:scoring_impl}). Automated prompt tuning is subsequently performed using the TextGrad optimization loop (Section \ref{subsec:opt_loop}) to refine the simulator behavior across tasks. 
\subsection{Bias Correction via Entropy-Based Diversity Metrics}
LLM simulators exhibit systematic biases that undermine evaluation validity: \textbf{positive bias} (over-acceptance of recommendations) \cite{Yoon2024LLMsAsUserSimulators}, \textbf{popularity bias} (over-representation of well-known items) \cite{Zhang2020EvalCRSUserSimulation,Zhao2021UserSim}, and \textbf{temporal clustering} (narrow time-window recommendations). Prompt sensitivity further amplifies these issues. To model behavioral diversity, we use entropy-based metrics where higher entropy reflects greater unpredictability. We define five complementary measures and each of them is designed to capture different aspects of human behavioral fidelity: (1) \textbf{Item Entropy} captures diversity across genre, era, region, and cultural origin, penalizing duplicates and IMDb Top 250 overuse; (2) \textbf{BinPref} quantifies alignment between simulator and human binary preferences; (3) \textbf{Aspect Entropy} measures justification diversity (e.g., plot, acting, pacing); (4) \textbf{Sentiment Entropy} reflects tone variation, exposing positive bias; (5) \textbf{Semantic Richness} assesses vocabulary diversity. Our method iteratively optimizes prompts using these metrics, comparing against ground truth patterns and generating feedback to reduce bias:

\begin{enumerate}
    \item \textbf{ItemsTalk}: Item entropy quantifying diversity in movie selections. Let $X = \{x_1, \dots, x_n\}$ be the list of movies binned by (genre, era, region):
    \begin{equation}
    H(X) = -\sum_{i=1}^{n} p(x_i) \log p(x_i) \label{eq:item_entropy}
    \end{equation}
    
    \item \textbf{BinPref}: Pearson correlation between simulated acceptance rates $(x_i)$ and human movie ratings $(y_i)$:
    \begin{equation}
    r = \frac{\sum_{i=1}^{n} (x_i - \bar{x})(y_i - \bar{y})}{\sqrt{\sum_{i=1}^{n} (x_i - \bar{x})^2 \sum_{i=1}^{n} (y_i - \bar{y})^2}}
    \end{equation}
    
    \item \textbf{OpenPref}: Aspect and sentiment entropy measuring review richness. Let $X = \{x_1, \dots, x_n\}$ be aspects (plot, acting, pacing) and $Y = \{y_1, \dots, y_m\}$ be sentiments (neg, neu, pos):
    \begin{align}
    H_{\text{aspect}}(X) &= -\sum_{i=1}^{n} p(x_i) \log p(x_i) \\
    H_{\text{sentiment}}(Y) &= -\sum_{i=1}^{m} p(y_i) \log p(y_i)
    \end{align}
    
    \item \textbf{RecRequest}: Multiple diversity measures. Let $\mathbf{w}_i$ be word vectors and $\mathbf{s}_i$ be sentence vectors:
    \begin{align}
    \text{TTR} &= \frac{|\text{Types}|}{|\text{Tokens}|} \\
    \text{W2V} &= 1 - \frac{1}{|\text{Words}|^2} \sum_{i,j=1}^{|\text{Words}|} \cos(\mathbf{w}_i, \mathbf{w}_j) \\
    \text{SentDiv} &= 1 - \frac{1}{|\text{Sentences}|^2} \sum_{i,j=1}^{|\text{Sentences}|} \cos(\mathbf{s}_i, \mathbf{s}_j)
    \end{align}
\end{enumerate}

A potential concern with entropy-based optimization is that maximizing entropy could produce outputs that are diverse but behaviorally random rather than human-like. Our framework addresses this directly by anchoring all scoring functions to empirically derived human baseline distributions rather than rewarding raw entropy maximization. Concretely, our evaluation prompts define target entropy ranges derived from real Amazon user behavior. Outputs that fall outside these ranges in either direction are penalized, meaning the optimizer is explicitly discouraged from overshooting into unrealistic diversity. 
\subsection{Profile Summarization for Data Leakage Prevention}
\label{subsec:profile_summary_for_data}
Using complete user histories creates context overload and data leakage when target items appear in history. We introduce profile summarization: randomly sample user history subsets and abstract into natural language preferences, e.g., transforming \emph{The Godfather, Goodfellas, Casino} into ``prefers character-driven crime dramas with moral ambiguity." 
For each user, we sample up to $k=5$ items from their history, and pass titles and years to Llama-3.3-70B with the following constrained prompt:

\begin{quote}
\textit{``Summarize the user's taste and behavior in simple English. Keep the description short without naming specific movies or TV shows.''}
\end{quote}
Refer to section \ref{subsec:profile_generation} for results.

\subsection{NegFeedback: Rejection Quality Evaluation}
\label{subsec:negfeedback}
We designed a new metric, NegFeedback, extending \cite{Yoon2024LLMsAsUserSimulators}'s feedback task to assess rejection quality. Unlike prior work focused on accept/reject coherence, NegFeedback evaluates if simulators give persona-aligned rejection rationales and informative reformulation requests, revealing deeper reasoning capability. NegFeedback requires simulators to: reject inappropriate recommendations, provide preference based justifications, generate semantically rich reformulation requests without naming movies, and maintain persona-consistent tone. An LLM-based evaluator scores responses (0--10) on rejection clarity, justification specificity, reformulation quality, and behavioral consistency. For each NegFeedback instance, the evaluator LLM (Llama-3.3-70B) receives the user profile summary, recommended item description, and simulator response, along with a rubric describing four criteria: (A) \emph{rejection clarity}, (B) \emph{persona-grounded justification specificity}, (C) \emph{reformulation quality without naming specific titles}, and (D) \emph{tone consistency with persona}. The evaluator outputs four integer scores $s_A, s_B, s_C, s_D \in [0,10]$, and we define
\begin{equation}
\text{NegFeedback} = \frac{1}{4} (s_A + s_B + s_C + s_D).
\end{equation}
 The same rubric is used for human evaluation and we report inter-annotator agreement and human--LLM correlation in section \ref{subsec:negfeedback_analysis}. 
\subsection{TextGrad Optimization Loop}
\label{subsec:opt_loop}

We apply TextGrad optimization \textbf{separately for each task}, allowing task-specific prompt refinement. For task $i$, we:

\begin{enumerate}
    \item Initialize prompt $\theta^{(0)}_i$ with a task-specific template
    \item For iteration $t = 0, \dots, T-1$:
    \begin{enumerate}
        \item[(a)] Sample batch of users $B$, generate responses using $\theta^{(t)}_i$
        \item[(b)] Compute task-specific metric $m_i(\theta^{(t)}_i)$ 
        \item[(c)] Query back engine LLM in Textgrad for feedback on improving $m_i$
        \item[(d)] Update prompt to $\theta^{(t+1)}_i$
    \end{enumerate}
    \item Use optimized $\theta^{(T)}_i$ for final evaluation
\end{enumerate}
In all experiments we run $T = 3$ iterations with batch size $B = 100$ users. Optimization stops at the last iteration. Only the guideline and persona sections of the prompt are editable; task instructions and system boilerplate remain fixed. Edits are constrained to preserve field structure and keep total prompt length below $L = 2048$ tokens.
\subsection{Text Gradient Design}
\label{subsec:scoring_impl}
\paragraph{Textgrad Evaluation Prompt Design}
For each task, we first established human baselines using responses from 10 randomly selected Amazon users. These baselines defined the target distributions for the five tasks. Using these values, we constructed an initial evaluation prompt and performed prompt tuning. After each iteration, the resulting textual gradients were manually analyzed to extract qualitative feedback in the form of natural language. Insights from this analysis were then used to refine and update the evaluation prompt by incorporating additional constraints, ensuring that the evaluation prompt progressively and accurately captured and evaluated the intended aspects of each task. Evaluation prompts were iteratively refined over multiple optimization runs using textual-gradient analysis until the simulator outputs aligned closely with the human targets. This refinement was performed separately for each task and concluded once the evaluation prompt stabilized. The finalized prompts were then used for large-scale evaluation. Once the final evaluation prompt was established, we scaled the evaluation by replacing textual-gradient analysis with numerical scoring functions. These functions mapped deviations from the human baseline to scores in the range [1,10], where 1 indicates the largest deviation and 10 indicates the closest match.

\begin{table*}[htbp]
\caption{Prompt Evolution Across Five Tasks Optimization}
\label{tab:prompt_examples_all}
\centering
\footnotesize
\begin{tabular}{p{0.15\textwidth}|p{0.38\textwidth}|p{0.38\textwidth}}
\hline
\textbf{Task} & \textbf{Before Optimization (Vanilla)} & \textbf{After Optimization} \\
\hline\hline

\textbf{ItemsTalk} \newline \textit{(Diversity)} 
& 
\textit{``Pretend to be Mr Zaroui (MENA), Age: 60, Pickiness: moderately picky. Profile: likes a mix of old and new movies from 1988--2022. Talk about 4 movies. Reply as "Title (yyyy)". Say nothing else.''} 
& 
\textit{``Pretend to be Mr. Zaroui (MENA), Age: 60, film enthusiast favoring \textbf{international cinema, independent films, documentaries}. Provide 4 titles ensuring: \textbf{(1)} $\geq$4 continents, \textbf{(2)} temporal diversity (pre-1980, 90s, 2000s, post-2010), \textbf{(3)} max 1 film per genre, \textbf{(4)} award-winning but lesser-known, \textbf{(5)} no horror. Maximize item entropy across region/genre/era.''} \\
\hline
\multicolumn{3}{p{0.96\textwidth}}{\textit{Impact:} Item entropy: 6.4 $\rightarrow$ 9.3 (+44.6\%)} \\
\hline

\textbf{BinPref} \newline \textit{(Preference Alignment)} 
& 
\textit{``Pretend to be Ms Sethi (South Asian), Age: 48, Pickiness: not picky. You watched the movie Bodies Bodies Bodies (2022) and rated it 3.43 out of 5. Did you like the movie? Answer Yes or No. Don't say anything else.''} 
& 
\textit{``Pretend to be Ms. Sethi, a 48-year-old woman from India, who watched "Bodies Bodies Bodies" (2022) and rated it 3.43/5. \textbf{Consider a movie liked if rating $\geq$3.5}. Output \textbf{1} if rating $\geq$3.5 (liked), \textbf{0} otherwise, focusing solely on the provided rating.''} \\
\hline
\multicolumn{3}{p{0.96\textwidth}}{\textit{Impact:} Pearson correlation: 0.265 (GPT-4) $\rightarrow$ 0.726 (+174\%)} \\
\hline

\textbf{OpenPref} \newline \textit{(Review Richness)} 
& 
\textit{``Pretend to be Ms Park (East Asian), Age: 8, Pickiness: not picky. You watched the movie Seven Girlfriends (2000). What are your thoughts on this movie?''} 
& 
\textit{``Pretend to be Ms. Park, an 8-year-old East Asian girl who loves movies. You watched Seven Girlfriends (2000), a romantic comedy. Focusing on \textbf{relationships}, provide $\geq$2 detailed examples of what you liked/disliked about this theme, considering storyline, color use, character development. Describe a specific scene portraying relationship complexity and explain why it stood out. How does Jesse's journey resonate with your \textbf{own friendships/personal growth}? Use transitional phrases, varied sentence structures, and analyze one theme deeply.''} \\
\hline
\multicolumn{3}{p{0.96\textwidth}}{\textit{Impact:} Aspect entropy: 5.087 (GPT-4) $\rightarrow$ 7.523 (+47.9\%); Sentiment entropy: 0.000 $\rightarrow$ 1.274} \\
\hline

\textbf{RecRequest} \newline \textit{(Semantic Richness)} 
& 
\textit{``Generate a movie recommendation request. Include (but do not request) the following movies in your text: ['Last Train Home (English Subtitled) (2010)', 'The Secret File of Marco Polo (2021)'].''} 
& 
\textit{``Recommend a \textbf{psychologically complex suspenseful thriller} with \textbf{strong emotional resonance, unexpected twists}, and a \textbf{female lead}.''} \\
\hline
\multicolumn{3}{p{0.96\textwidth}}{\textit{Impact:} TTR: 0.552 (GPT-3.5) $\rightarrow$ 0.235 (closer to human 0.188); W2V diversity: 0.471 $\rightarrow$ 0.779 (+65.4\%)} \\
\hline
\textbf{NegFeedback} \newline \textit{(Rejection Quality)} 
& 
\textit{``You are now role-playing as Ms Garcia (Mixed or Multiracial), Age: 19, Pickiness: not picky. Your preferences: likes classic TV shows from the past, especially action and detective series, as well as thought-provoking movies like psychological thrillers. Enjoys nostalgic content with a mix of entertainment and suspense. You received a movie recommendation: Genndy Tartakovsky's 'Primal' – Tales of Savagery (2020). You do not like this recommendation. Your task: 1. Clearly reject the movie. 2. Explain why it doesn't match your preferences (tone, genre, theme). 3. Formulate your request without naming any specific movie. 4. Match your tone and reasoning style. 5. If unable, respond: "I don't know."''} 
& 
\textit{``You are Ms. Garcia (Mixed/Multiracial), Age: 19, who loves \textbf{classic detective shows with clever puzzle-solving} (e.g., Columbo's methodical investigations) and \textbf{psychological thrillers with layered mysteries}. You received: Genndy Tartakovsky's 'Primal' – Tales of Savagery (2020). \textbf{Reject it clearly}, explaining why its tone/genre/theme clashes with your preferences for \textbf{dialogue-driven mysteries, intelligent female leads, and sophisticated suspense}. Then reformulate your request using \textbf{abstract descriptors}: desire films blending atmospheric tension with intricate plot twists, vintage-inspired aesthetics (1950s-60s), strong character dynamics, and cerebral crime-solving—\textbf{without naming any movies}. Match a casual, witty tone reflecting your love of classic Bond-style sophistication. Keep response under 1200 characters.''} \\
\hline
\multicolumn{3}{p{0.96\textwidth}}{\textit{Impact:} LLM-human correlation: -0.244 (GPT-3.5) $\rightarrow$ 0.484 (+198\%)} \\
\hline
\end{tabular}

\begin{minipage}{\textwidth}
\footnotesize
\textbf{Key Patterns observed after prompt tuning:} (1) \textit{ItemsTalk}: Adds explicit diversity constraints (geography, time, genre); (2) \textit{BinPref}: Introduces threshold rule (3.5 cutoff) and binary output format; (3) \textit{OpenPref}: Demands multi-aspect analysis, scene-specific examples, and personal reflection; (4) \textit{RecRequest}: Removes verbatim titles, adds rich semantic descriptors.(5) \textit{NegFeedback}: Enhances rejection quality,reformulation ability, and rationale coherence 
\end{minipage}
\end{table*}
\paragraph{Task-Specific Evaluation Prompts}
\label{subsec:eval_prompts}

Each task employs a specialized evaluation prompt $e_i$ that defines hard constraints, scoring criteria, and target behavioral ranges informed by human baselines. In \textbf{ItemsTalk}, $e_i$ enforces strict formatting before evaluating content, penalizing popular or redundant selections and rewarding high-entropy lists spanning various genres, decades, and continents. In \textbf{BinPref}, $e_i$ uses deterministic correctness scoring, assigning positive credit only to exact \texttt{Yes}/\texttt{No} outputs with no additional text, aligned with human decision boundaries. In \textbf{OpenPref}, $e_i$ constrains responses with respect to aspect, and sentiment entropy baselines derived from human review distributions. In \textbf{RecRequests}, $e_i$ assess fluency, implicitness, and diversity without explicit item references, targeting baselines based on human-written queries. In \textbf{NegFeedback}, $e_i$ emphasizes clear rejection justification tied to persona preferences, semantically rich reformulation without naming items, and tone consistency, with a 10-point rubric penalizing generic reasoning, verbatim movie mentions, and excessive prompt length. All tasks use fixed $K=3$ epochs without early stopping. Table~\ref{tab:prompt_examples_all} illustrates the concrete prompt transformations resulting from this optimization process. Table~\ref{tab:eval_prompts} presents the complete evaluation prompts used for optimization.

\begin{table*}[htbp]
\caption{Task-Specific Evaluation for Prompt Optimization}
\label{tab:eval_prompts}
\centering
\scriptsize
\begin{tabular}{p{0.12\textwidth}|p{0.83\textwidth}}
\hline
\textbf{Task} & \textbf{Evaluation Prompt ($e_i$)} \\
\hline\hline

\textbf{Task 1:} \newline \textit{ItemsTalk} \newline (Diversity)
& 
You are an evaluator. Assign a numeric score to a list of movie titles based on the following rules.

\textbf{HARD FORMAT CONSTRAINTS (must pass to score $>$2):}
\begin{itemize}
\item Each line: \texttt{Title (YYYY)} — no commentary, numbering, or bullets
\item No duplicate titles, no IMDb Top 250 movies, no popular blockbusters
\item $\leq$1 film per franchise/director (unless requested)
\end{itemize}
If violated, assign score 0--2 and STOP.

\textbf{CONTENT CRITERIA (only if format valid):}
Diversity Requirements:
\begin{itemize}
\item $\geq$5 distinct genres (documentary, animation, experimental, drama, sci-fi)
\item $\geq$3 different decades (e.g., 1970s, 1990s, 2010s)
\item $\geq$3 continents/cultural regions (Asia, Africa, South America)
\item Mix of styles: live-action, animated, documentary, silent, avant-garde
\item No two entries similar in theme, tone, or origin
\end{itemize}

Item entropy measures unpredictability and novelty. High entropy = avoids repetition, genre clustering, cultural bias. Must be $\geq$20 for top score.

\textbf{SCORING:} 0--2: Format violations; 3--4: Entropy $<$10; 5--7: Entropy 10--20; 8--10: Entropy $\geq$20.

\textbf{OUTPUT:} Single digit 0--10. No explanation. \\
\hline

\textbf{Task 2:} \newline \textit{BinPref} \newline (Preference Alignment)
& 
You are an automatic evaluator. Rules:
\begin{itemize}
\item Return \textbf{1} if output is \texttt{Yes} for high-rated movies (rated $>$3.5/5) or \texttt{No} otherwise, with no extra text
\item Return \textbf{0} if format incorrect, contains extra text, or misaligns with rating/preference
\end{itemize}

\textbf{OUTPUT:} Single digit 1 or 0. \\
\hline

\textbf{Task 3:} \newline \textit{OpenPref} \newline (Review Richness)
& 
\textbf{Aspect Taxonomy (choose exactly 1):} Story, Characters, Pacing, Visuals, Sound, Acting, Representation

\textbf{Evaluation Criteria:}
\begin{itemize}
\item Use exactly 1 aspect. Using $>$1 results in penalty.
\item Aspect entropy target: 5.0--8.0
\item Sentiment entropy target: 1.2--1.5
\item No vague language, jargon, or generic filler
\end{itemize}

\textbf{SCORING:}
\begin{itemize}
\item 0--2: Violates hard constraints ($>$1 aspect, entropy $>$$\pm$2 from target)
\item 3--4: Minor violations (2 aspects or entropy $>$$\pm$1.5)
\item 5--7: Meets aspect count; entropy slightly outside target ($>$$\pm$0.2); mostly relevant
\item 8--10: Meets aspect count; entropy within target; concise, specific, human-like
\end{itemize}

\textbf{OUTPUT:} Single digit 0--10. No extra text. \\
\hline

\textbf{Task 4:} \newline \textit{RecRequest} \newline (Semantic Richness)
& 
\textbf{Evaluation Criteria:}
\begin{itemize}
\item \textbf{Implicitness:} Do not name/list reference movies
\item \textbf{Natural tone:} Common words, smooth phrasing, no jargon
\item \textbf{Constraint variety:} Include 2--4 distinct cues (tone, genre, mood, setting)
\item \textbf{Fluency:} 3--6 well-structured sentences with natural flow
\item \textbf{Diversity Metrics:}
  \begin{itemize}
  \item Word diversity (TTR): 0.1--0.25
  \item Word embedding diversity (W2V): 0.35--0.55
  \item Sentence embedding diversity: 0.35--0.55
  \end{itemize}
\end{itemize}

\textbf{SCORING:}
\begin{itemize}
\item 0--2: Named references, poor fluency, diversity $>$$\pm$0.2 from target
\item 3--4: Minor issues; diversity $>$$\pm$0.1 from target
\item 5--7: All constraints met; diversity within $\pm$0.05; mostly fluent
\item 8--10: Fully human-like; all constraints met; strictly within range; fluent, implicit, expressive
\end{itemize}

\textbf{OUTPUT:} Single digit 0--10. No extra text. \\
\hline

\textbf{Task 5:} \newline \textit{NegFeedback} \newline (Rejection Quality)
& 
Evaluate the simulator response for:
\begin{itemize}
\item \textbf{A.} Clear rejection justified by persona's preferences (tone, genre, theme)
\item \textbf{B.} Semantically rich, persona-aligned reformulation — \textbf{without naming any movie}
\item \textbf{C.} Tone and reasoning style matching the persona
\end{itemize}

\textbf{SCORING RUBRIC:}
\begin{itemize}[]
\item 0: Responds ``I don't know'' when meaningful response possible
\item 1: No rejection + No reason + No reformulation OR mentions movies
\item 2--3: One weak/irrelevant element OR mentions movies
\item 4--6: Generic rejection + Weak/missing reformulation OR mentions movies
\item 7: Generic rejection + Vague reformulation + No movies + Length $>$2000 chars
\item 8: Mildly generic + Vague reformulation + Tone inconsistent + No movies + Length $<$2000 chars
\item 9: Mildly generic + Less precise reformulation + Tone slightly off + No movies + Length $<$1500 chars
\item 10: \textbf{Full compliance}: Clear rejection, specific justification tied to persona, rich reformulation, tone match, no movies named, length $<$1200 chars
\end{itemize}

\textbf{OUTPUT:} Single score 0--10. No explanation. \\
\hline
\end{tabular}

\begin{minipage}{\textwidth}
\footnotesize
\textbf{Note:} All prompts instruct the evaluator LLM (Llama-3.3-70B) to return only a numeric score (0--10 or 0--1 for binary tasks) with no additional text, ensuring deterministic optimization signals for automated prompt tuning. Target ranges for entropy and diversity metrics were empirically derived from human baseline distributions (Section~\ref{subsec:dataset}).
\end{minipage}
\end{table*}
\section{Experiments}
\label{sec:experiments}
Our experimental evaluation addresses three core research questions to validate the proposed framework. \textbf{RQ1: Does our framework improve performance over existing prompt-based user simulators in conversational recommender systems?} To answer this question, we compare our method against GPT-3.5-turbo and GPT-4 baselines, which represent the state-of-the-art prompt-based user simulation approaches used in prior research \cite{Yoon2024LLMsAsUserSimulators,zhu2024reliable,zhu2025llm}. Results are presented in Section~\ref{subsec: Table results}. \textbf{RQ2: How does our framework perform under different prompting configurations?} To answer this question, we conduct comprehensive comparisons across four settings: zero-shot, In-Context Learning (ICL), Retrieval-Augmented Generation (RAG), and Profile-Augmented Generation (PAG), following established evaluation protocols \cite{zhuang2024hydra}. Comparative results across these configurations are reported in Section~\ref{subsec: Table results}. \textbf{RQ3: Is the proposed NegFeedback metric reliable for measuring simulators' rejection quality and mitigating over-acceptance bias?} To answer this question, we recognize that while human evaluation provides the gold standard for assessing rejection behavior \cite{Yoon2024LLMsAsUserSimulators,Sun2021SimulatingUserSatisfaction}, it does not scale to large-scale evaluation \cite{zhu2024reliable,zhang2024usimagent}. Therefore, we design an LLM-based evaluator to enable scalable assessment. To guarantee the quality of the LLM evaluator, we conduct a pilot human evaluation study and calculate the alignment between human judgments and LLM evaluator scores using Pearson correlation. The human validation methodology is detailed in Section~\ref{subsec:negfeedback_human_baseline}, with comprehensive NegFeedback performance analysis presented in Section~\ref{subsec:negfeedback_analysis}.
\subsection{Dataset}
\label{subsec:dataset}
We use the Amazon Reviews 2023 dataset (Movies \& TV subset) \cite{hou2024bridging}, which provides large scale real world user interactions ($>$1.4 million reviews) with diverse behavioral patterns. 
We develop a pre-processing pipeline that transforms raw review data into structured formats suitable for our five simulation tasks. The Movies \& TV subset aligns with conversational recommendation research and provides comparable baselines from prior work. The methodology generalizes across other Amazon Reviews 2023 corpus subsets, ensuring scalability and adaptability for future research.

\subsection{Deployment Infrastructure and Practical Considerations}
\label{subsec:deployment}
We implemented a custom \texttt{OllamaEngine} extending TextGrad's \texttt{EngineLM} for local deployment using Llama 3.3:70B, selected for its instruction-following capability and compatibility with privacy-preserving execution. Smaller models (e.g., mistral-small, deepseek-r1) failed to meet TextGrad’s requirements. Our system ran on 1 NVIDIA A100 GPU via institutional HPC, completing simulation and evaluation for 100 users across 5 tasks in 15 GPU hours. Profiles are generated independently, and optimized prompts are reusable, enabling low-cost, parallelized evaluation. Cross-domain adaptation requires minimal changes to attribute schemas and prompt templates, demonstrating feasibility for organizations with diverse recommendation systems.

\subsection{Profile Generation and Summarization}
\label{subsec:profile_generation}

We implement a privacy-preserving profile generation system using locally deployed Llama 3.3:70B via Ollama, addressing cloud-related data concerns while maintaining generation quality. Profiles include demographic and behavioral traits based on prior work \cite{Zhang2025LLMPoweredUserSimulator}. Each profile is generated through structured LLM interaction to ensure diversity and coherence. To reduce the potential for direct data leakage, we summarize user history by sampling up to five movies and prompting the LLM to produce a concise preference description. This two-stage process  yields compact, natural language profiles suitable for simulation while anonymizing sensitive data \cite{zhu2024reliable,zhu2025llm,D2K}. The profile summarization output replaces the raw history in all initial prompts. This approach reduces length significantly while \emph{reducing the potential for direct data leakage, improving item entropy and substantially lowering context window pressure}. To assess leakage risk and diversity impact, we analyzed item entropy across 100 users: raw histories yielded simulator item entropy of 6.415, while summaries increased this to 9.269 (vs human baseline 10.067). We further analyzed 100 responses where movie names appeared in user history: raw histories contained target titles verbatim in almost all of the cases which resulted in the high predictability of the titles mentioned and led to low entropy score (6.415). This is strong evidence that data leakage caused the reduction of item entropy values. The profile summaries abstracted these movie titles to genre/preference descriptions in all of the cases. This was verified by manual inspection of 100 samples and found no verbatim title reproduction. The resultant entropy measured in this case was the closest to the human baseline. This diagnostic strongly validates profile summarization reduces direct leakage exposure and substantially improves behavioral diversity.

\subsection{Task Design and Baselines}
\label{subsec:task_design}
We evaluate on five tasks from \cite{Yoon2024LLMsAsUserSimulators}: \textbf{ItemsTalk} measures item mention diversity through entropy;  \textbf{BinPref} measures the preference alignment between user simulator and the real user; \textbf{OpenPref} assesses nuanced aspect and sentiment expression; \textbf{RecRequest} evaluates lexical diversity and request granularity. We compare against GPT-3.5-turbo and GPT-4 in four configurations: zero-shot, In-Context Learning (ICL), Retrieval Augmented Generation (RAG), and Profile Augmented Generation (PAG), following previous research\cite{zhuang2024hydra}. These baselines represent user simulators without heavy model training. For NegFeedback, we conduct human evaluation with independent evaluators assessing rejection clarity, behavioral consistency, and reformulation quality on 10 randomly sampled responses. We analyze inter-evaluator correlation and measure human evaluator agreement with LLM-based scores across our method and baseline models.

\subsection{NegFeedback Human and Baseline Evaluation}
\label{subsec:negfeedback_human_baseline}
We introduce \textbf{NegFeedback}, which evaluates rejection quality, reformulation ability, and rationale coherence which are core behaviors in conversational recommendation.
\textbf{Study Design:} We conducted a pilot human evaluation on 10 samples to establish proof-of-concept validity. Three native English-speaking graduate students independently scored LLM-generated rejection responses on a 0--10 scale across four criteria: \textbf{(A)} Rejection clarity (0=implicit acceptance, 10=unambiguous rejection), \textbf{(B)} Justification specificity (0=generic, 10=persona-grounded), \textbf{(C)} Reformulation quality (0=no guidance, 10=semantically rich without naming titles), \textbf{(D)} Tone consistency (0=contradictory, 10=perfect match). While $n=10$ is insufficient for definitive conclusions, results provide initial evidence of metric validity; we commit to scaling to $n\geq100$ in future work. \textbf{Protocol:} Evaluators were blinded to model identity, received randomized samples, and scored based on user profile + movie name + simulator rejection response. The same rubric was used for both human and LLM evaluation to reduce subjectivity and LLM bias. Inter-rater agreement among three evaluators shows moderate consistency (average pairwise correlation: $0.52$)

\subsection{Results and Discussion}

\subsubsection{Performance Across Behavioral Tasks}
\label{subsec: Table results} 
\begin{table}[htbp]
\caption{OpenPref Task Results}
\label{tab:openpref}
\centering
\begin{tabular}{l c c}
\hline
\textbf{Model} & \textbf{Aspect Entropy} & \textbf{Sent Entropy} \\
\hline
Human                                   & 6.761 & 1.277 \\
GPT-3.5-turbo                    & 4.322 & 0.000 \\
GPT-3.5-turbo(ICL)      & 4.524 & 0.000 \\
GPT-3.5-turbo(RAG)             & 4.170 & 0.000 \\
GPT-3.5-turbo(PAG)            & 4.248 & 0.000 \\
GPT-4                    & 5.087 & 0.000 \\
GPT-4 (RAG)             & 5.087 & 0.000 \\
\textbf{llama3.3:70b}       & \textbf{4.720} & \textbf{1.363} \\
\textbf{Our Method}       & \textbf{7.523} & \textbf{1.274} \\
\hline
\end{tabular}
\end{table}

Our simulator demonstrates substantial improvements across multiple behavioral dimensions through automated prompt optimization. Apart from comparing with GPT baseline and our method, we include the human baseline for all the tasks in the tables. \textbf{Human} baseline is computed directly from Amazon user data, serving as the gold standard for behavioral fidelity. Unlike synthetic responses from simulators, these reflect genuine diversity, preferences, and communication patterns. 

In the OpenPref task (Table \ref{tab:openpref}), \textbf{post-tuning aspect entropy (7.523) approximates human performance (6.761) and dramatically exceeds GPT baselines (4.17 to 5.087), while maintaining realistic sentiment entropy (1.274) that matches human patterns (1.277).} GPT variants uniformly collapse to zero sentiment entropy across all context strategies (ICL, RAG, PAG), revealing severe positive bias. This failure stems from GPT's tendency to generate safe, uniformly positive responses when conditioned on user preferences, a bias that our entropy based optimization explicitly addresses through iterative feedback that penalizes sentiment uniformity.

\begin{table}
\caption{ItemsTalk Task Results}
\label{tab:itementropy}
\centering
\begin{tabular}{l c}
\hline
\textbf{Model} & \textbf{Item Entropy} \\
\hline
Human                            & 10.067 \\
GPT-3.5-turbo           & 6.644 \\
GPT-3.5-turbo (ICL)     & 6.596 \\
GPT-3.5-turbo (BM25 RAG)& 6.644 \\
GPT-3.5-turbo (PAG)     & 6.496 \\
GPT-4           & 6.596 \\
GPT-4 (BM25 RAG)& 6.496 \\
\textbf{llama3.3:70b}   & \textbf{7.626} \\
\textbf{Our Method}     & \textbf{9.269} \\
\hline
\end{tabular}
\end{table}

For ItemsTalk (Table \ref{tab:itementropy}), \textbf{post optimization achieves item entropy (9.269), closely matching human diversity (10.067) and substantially exceeding GPT baselines (6.496 to 6.644)}. The 9.269 item entropy (vs human 10.067) reflects improved diversity enabled by shorter, leakage-reduced summaries that prevent over-reliance on history-popular items. The uniform collapse across ICL, RAG, and PAG variants suggests these context augmentation strategies introduce information that overwhelms the GPT model, causing it to treat the augmented context as noise that degrades GPT's inherent diversity, with even vanilla GPT-3.5-turbo underperforming base Llama-3.3-70b likely due to GPT's extensive training constraining item selection toward safer mainstream recommendations and its training cutoff underrepresenting niche item diversity, both biasing toward high-frequency items. Our iterative optimization successfully amplifies the LLM diversity by the multiple iterations of tuning and reward based learning. 

\begin{table}[htbp]
\caption{BinPref Task Results}
\label{tab:binpref}
\centering
\begin{tabular}{l c}
\hline
\textbf{Model} & \textbf{Pearson Coeff} \\
\hline
Human & 1.000 \\
GPT-3.5-turbo           & 0.079 \\
GPT-3.5-turbo (ICL) &  0.259 \\
GPT-3.5-turbo (RAG)     &  0.265 \\
GPT-3.5-turbo (PAG)     & 0.234 \\
GPT-4          &  0.265 \\
GPT-4 (RAG)     &  0.265 \\
\textbf{llama3.3:70b}     & \textbf{0.731} \\
\textbf{Our Method}     & \textbf{0.726} \\
\hline
\end{tabular}
\end{table}

In BinPref(Table \ref{tab:binpref}), the models based on GPT-3.5-turbo and GPT-4 demonstrate relatively low correlation. Here \textbf{llama3.3:70b base (0.731) slightly outperforms our tuned method (0.726)}. This may be due to the fact that without the tuning, the base model had strong instruction following ability and prompt tuning in this case was not necessary. But prompt tuning performs better than GPT baselines as the instruction following ability of these models were low. This same problem had been identified \cite{Yoon2024LLMsAsUserSimulators} in previous research in the GPT family. Using a larger and better instruction following model seems to have solved the problem.

\begin{table}[htbp]
\caption{RecRequest Task Results}
\label{tab:recrequest}
\centering
\begin{tabular}{l c c c}
\hline
\textbf{Model} & \textbf{Word Div} & \textbf{Word2Vec Div} & \textbf{Sentence Div} \\
\hline
Human                                  & 0.188 & 0.581 & 0.448\\
GPT-3.5-turbo                          & 0.552 & 0.471 & 0.542\\
GPT-3.5-turbo(ICL)            & 0.595 & 0.341 & 0.463 \\
GPT-3.5-turbo(RAG)                    & 0.597 & 0.431 & 0.530 \\
GPT-3.5-turbo(PAG)                    & 0.595 & 0.406 & 0.550 \\
GPT4                          & 0.421 & 0.404 & 0.535 \\
GPT4(RAG)                    &  0.437 &  0.382 & 0.545 \\
\textbf{llama3.3:70b}         & \textbf{0.394} & \textbf{0.605} & \textbf{0.571} \\
\textbf{Our Method}      & \textbf{0.235} & \textbf{0.779} & \textbf{0.609} \\
\hline
\end{tabular}
\end{table}

In RecRequest(Table \ref{tab:recrequest}), \textbf{post tuning achieves Type-Token Ratio of (0.235), closely approximating human lexical patterns (0.188), while GPT baselines exhibit inflated diversity (0.42+). }This overestimation indicates that GPT generates vocabulary rich but stylistically unnatural requests while prioritizing lexical variety over conversational authenticity. Our method's human baseline optimization successfully calibrates this trade off, producing requests that are appropriately repetitive in structure while remaining semantically meaningful.

\subsubsection{NegFeedback: Evaluating Rejection Realism}
\label{subsec:negfeedback_analysis}
\begin{table}
\caption{NegFeedback Task Results}
\label{tab:negfeedback}
\centering
\begin{tabular}{l c c} 
\hline
\textbf{Model} & \textbf{Corr with H.E} & \textbf{p-value}  \\
\hline
GPT-3.5-turbo                          & -0.2443 & 0.4964 \\
GPT-3.5-turbo(ICL)                     & -0.2137 & 0.5532 \\
GPT-3.5-turbo(PAG)                     & -0.4232 & 0.2230 \\
GPT-3.5-turbo(RAG)                     & 0.2584  & 0.4710 \\
\textbf{llama3.3:70b}                & \textbf{0.2584} & \textbf{0.4710 } \\
\textbf{Our Method}                  & \textbf{0.4840 }& \textbf{0.1564} \\
\hline
\end{tabular}
\end{table}
Table~\ref{tab:negfeedback} presents correlations between model outputs and human evaluation for the NegFeedback task. As detailed in Section~\ref{subsec:negfeedback_human_baseline}, our method achieves the highest human-LLM correlation ($r=0.484$, $p=0.156$), substantially outperforming base Llama ($r=0.258$) and all GPT-3.5-turbo variants (negative or near-zero correlations). Most GPT configurations produced zero-variance outputs, yielding undefined correlations which were corrected after Laplace smoothing. Our method achieves highest human-LLM correlation ($r=0.484$, $p=0.156$) vs. base Llama ($r=0.258$) and GPT-3.5-turbo variants (negative correlations, Table~\ref{tab:negfeedback}). While not statistically significant due to limited sample size, the positive moderate correlation suggests the LLM evaluator captures meaningful human-aligned signal. GPT-3.5-turbo configurations produced near-zero variance outputs, resulting in undefined correlations which needed Laplace smoothing. Yet these configurations gave poor correlations.

\section{Conclusion}
We presented a multi-objective framework for automatic prompt optimization of LLM-based user simulators in conversational recommender systems. This work jointly targets systematic positive bias, data leakage, and constrained behavioral diversity while avoiding brittle manual prompt engineering via textual gradients and entropy-aware scoring. Profile summarization is integrated to abstract verbatim histories. Our approach substantially optimizes item entropy, aspect and sentiment entropy, preference alignment, semantic richness, and rejection quality with respect to human baselines. The proposed NegFeedback task and metric further capture persona-grounded rejection quality and over-acceptance bias, with an LLM evaluator whose scores show meaningful correlation with human judgments, suggesting that scalable, human-aligned evaluation of rejection behavior is feasible without exhaustive human annotation. The approach generalizes across recommendation domains with minimal adaptation,enabling scalable evaluation for industrial use. This makes our approach immediately applicable for practitioners seeking reproducible, privacy compliant simulation pipelines in regulated or resource-constrained environments.

\section{AI-Generated Content Acknowledgement}
The authors acknowledge the use of Claude (Anthropic) to refine the language and improve the clarity of this manuscript. Specifically, the tool was used in the Introduction, Our Framework, and Experiments sections to polish author written text for better flow and grammatical precision. All AI-generated suggestions were critically reviewed and modified by the authors, who maintain full responsibility for the accuracy and integrity of the final content.

\bibliographystyle{IEEEtran}
\bibliography{custom}
\end{document}